# Past Achievements and Future Challenges in 3D Photonic Metamaterials

Costas M. Soukoulis*
Ames Laboratory and Department of Physics and Astronomy, Iowa State University, Ames, Iowa, 50011, USA
& Institute of Electronic Structure and Laser (IESL), FORTH 71110 Heraklion, Crete, Greece.

Martin Wegener
Institute of Applied Physics, Institute of Nanotechnology, and DFG-Center for Functional Nanostructures (CFN) at Karlsruhe Institute of Technology (KIT), D-76128 Karlsruhe, Germany.

\* soukoulis@ameslab.gov

*Abstract*

Photonic metamaterials are man-made structures composed of tailored micro- or nanostructured metallo-dielectric sub-wavelength building blocks that are densely packed into an effective material. This deceptively simple, yet powerful, truly revolutionary concept allows for achieving novel, unusual, and sometimes even unheard-of optical properties, such as magnetism at optical frequencies, negative refractive indices, large positive refractive indices, zero reflection *via* impedance matching, perfect absorption, giant circular dichroism, or enhanced nonlinear optical properties. Possible applications of metamaterials comprise ultrahigh-resolution imaging systems, compact polarization optics, and cloaking devices. This review describes the experimental progress recently made fabricating three-dimensional metamaterial structures and discusses some remaining future challenges.

## Introduction

Metamaterials, introduced about a decade ago[1-5], represents a broad class of micro- or nanostructures composed of tailored building blocks ideally much smaller than the wavelength of light, enabling dense packing into an effective material. Mie-like resonances of the building blocks are the key to open the door to a new world of possibilities. In particular, magnetic-dipole resonances can be achieved by "simply" shrinking the size of usual macroscopic metallic electromagnets such that their resonances enter the optical regime. The split-ring resonator (SRR) continues to be a paradigm example[6]. The incident light field can induce a circulating and oscillating electric current in the ring, leading to a magnetic dipole-moment normal to the ring. Naively, the SRR can be viewed as a half-wave antenna rolled into an almost closed circle. This picture leads to a resonance wavelength that is $2\pi$ times larger than the ring's diameter—not too far off more detailed calculations that also must account for plasmonic effects that occur when the operation frequency, even remotely, approaches the metals plasma frequency, gradually turning the antenna resonance into a Mie resonance. The magnetic-dipole resonance of the SRR and its numerous variations have reminded the optics community that light is an electromagnetic wave. If one aims at obtaining complete control over an electromagnetic light wave inside a material, one needs to be able to independently control both the electric and the magnetic fields[1,2]. Thus, the newly achieved magnetic control has opened one half-space of optics still considered irrelevant in many optics textbooks. Yet, it shouldn't. The other half of optics allows for impedance matching; hence, zero reflection at a material's interface for any refractive index. The refractive index can also become negative, which means that the phase velocity of light is opposite to the electromagnetic energy flow described by the Poynting vector (precisely, the dot product of the two vectors is negative). In other structures, the refractive index can assume exceptionally large positive values, *i.e.*, the phase velocity is very small. Huge optical activity arises if the refractive index is largely different for left- and right-handed circular polarization of light. Finally, to realize the concepts of transformation optics, one also needs tailored and generally anisotropic magneto-dielectric metamaterials.

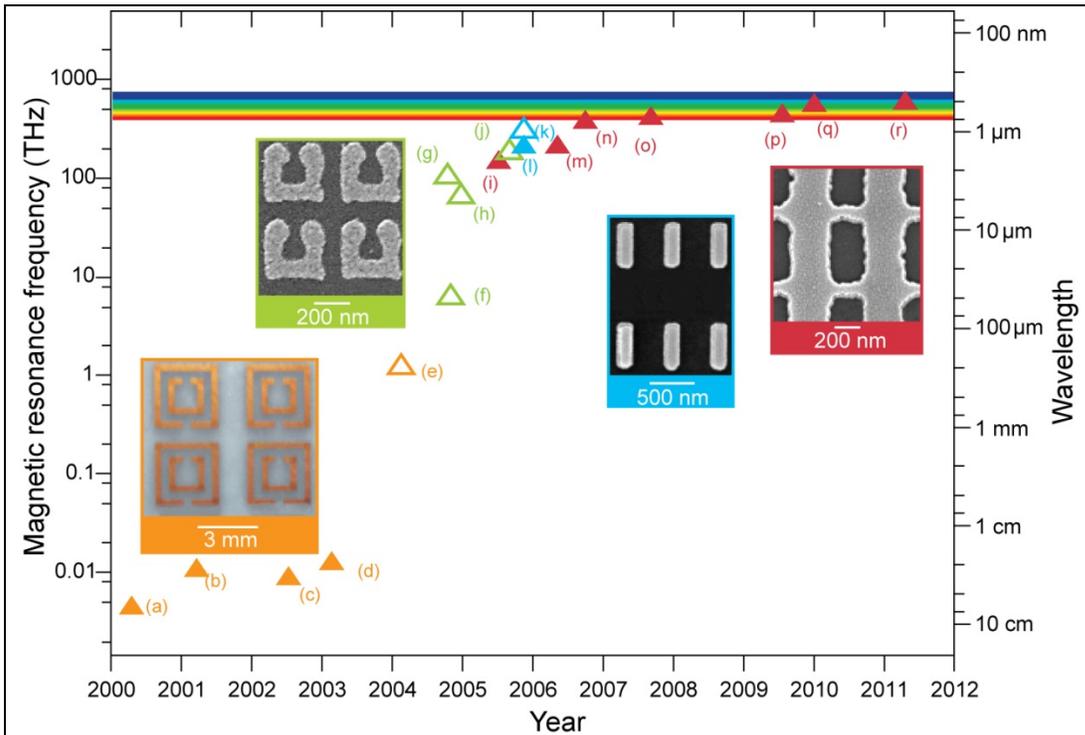

**Figure 1 Metamaterial operation frequency versus year.** The operation frequency of metamaterials with negative magnetic permeability $\mu$ (open symbols) and negative index of refraction $n$ (solid symbols) is shown on a logarithmic scale ranging from microwaves to the visible versus year over the course of the last decade. Orange: data from structures based on double split-ring resonators (SRR); green: data from U-shaped SRR; blue: data from metallic cut-wire pairs; red: data from negative-index double-fishnet structures. The four insets illustrate the designs by optical (only orange) or electron micrographs. (a)=[7], (b)=[8], (c)=[9], (d) =[10], (e) =[11], (f) =[12], (g) =[13], (h) =[14], (i) =[15], (j) =[16], (k) =[17], (l) =[18], (m) =[19], (n) =[20], (o) =[21], (p) =[22], (q) =[23], and (r)=[24].

In brief, there has never been a shortage of new ideas and dreams in the field of metamaterials. However, to make the transition from a curious scientific finding to a real-world material usable by the optics industry, several conditions must be fulfilled. First, the operation frequency must be brought to the optical, including telecom or visible (although there may be applications at longer wavelengths as well). As we will briefly recall below, this condition was met a few years ago. Second, to qualify as a "material," the structures should have several or even many layers of truly three-dimensional unit cells to approach the bulk 3D limit. We focus on this aspect within the present review. Third, the losses (absorption) should be reasonably low. Experimental progress along this direction has been sluggish for metal-based metamaterials[3-5], but all-dielectric structures avoiding metals promise a remedy in some frequency regimes.

*Metamaterials go optical*

In 2007, on the occasion of early visible negative-index metamaterials, we reviewed[2] the development of the metamaterial operation frequency for magnetic and/or negative-index metamaterials versus year. Figure 1 is an updated version of the graphics shown there. This tremendous increase of the operation frequency has become possible by miniaturizing and redesigning the magnetic SRRs. In contrast, the negative electric permittivity simultaneously required for negative refractive indices is achieved by sets of long metal wires ("diluted metal"), both at microwave and at visible frequencies. This redesign process has led to the so-called double-fishnet negative-index structure[15], which has been used in experiments by various groups[15, 17-24]. In essence, one functional layer of this structure is a perforated metal-dielectric-metal sandwich. The uniaxial double-fishnet exhibits an effective negative index of refraction for propagation of light normal to its layers.

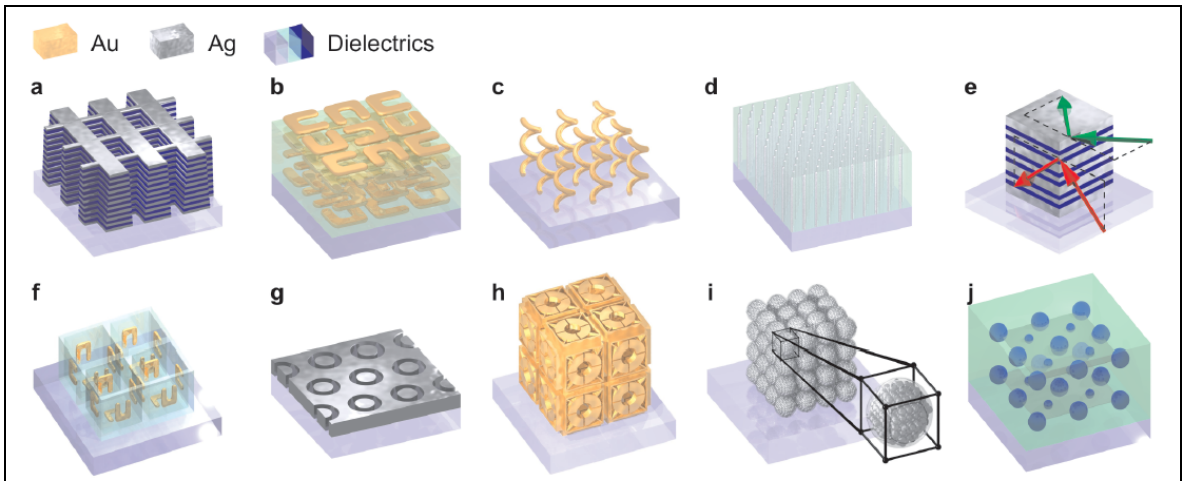

**Figure 2 Three-dimensional photonic-metamaterial structures.** (a) Double-fishnet negative-index metamaterials with several layers[22,24,28,30-32], (b) "stereo-" or chiral metamaterials (see also Fig. 3) *via* stacked electron-beam lithography[12,35-39,44-45], (c) chiral metamaterials made *via* direct-laser writing and electroplating[46], (d) hyperbolic (or indefinite) metamaterial[51,53-58] that can be made by electroplating of hexagonal-hole-array templates[57], (e) metal-dielectric layered metamaterial composed of coupled plasmonic waveguides enabling angle-independent negative refraction for particular frequencies[52,53], (f) split-ring resonators oriented in all three dimensions accessible *via* membrane projection lithography[59], (g) wide-angle visible negative-index metamaterial based on a coaxial design[97], (h) blueprint for a connected cubic-symmetry negative-index metamaterial structure amenable to direct laser writing[60], (i) blueprint for a metal cluster-of-clusters visible-frequency magnetic metamaterial amenable to large-area self-assembly[61], and (j) blueprint for an all-dielectric negative-index metamaterial composed of two sets of high-refractive-index dielectric spheres arranged on a simple-cubic lattice[62-65,75].

How can one provide experimental evidence that the index of refraction, $n$, is negative indeed? At optical frequencies, most groups have merely measured the frequency dependence of the transmittance, $T$, and/or of the reflectance, $R$, and have used computer simulations to fit the experimental values of $T$ and $R$. Next, the numerical results have been used to derive or "retrieve"[25] the effective parameters, *i.e.*, electric permittivity, $\varepsilon$, magnetic permeability, $\mu$, and refractive index, $n$ (and, more generally, also of the bi-anisotropy parameter[26,27]). This simple approach can still deliver reliable results, provided the agreement between experiment and theory is excellent. Yet, this approach is quite indirect and does not necessarily lead to unique results. For a single layer of fishnet structure, the unit cell size along the propagation direction, $a_z$, is undefined. Although the magnitude of $T$, and $R$ does not change, their phases change upon varying $a_z$. As $a_z$ decreases, the magnitude of the retrieved effective parameters tends to increase. In contrast, for two or more fishnet layers, the axial lattice constant $a_z$ is well defined, and the retrieved results only depend on the number of layers. If the retrieved parameters converge with increasing number of layers, "bulk" metamaterial properties are reached. (see discussion below). More directly, a negative $n$ should lead to (i) negative refraction at an interface and (ii) to a negative phase velocity of light; hence, to a negative propagation time of the phase fronts upon propagation through a metamaterial slab. Aspect (i) has been shown experimentally by using a wedge prism at telecom wavelengths[28], in analogy to corresponding experiments at microwave frequencies[8]. However, caution must generally be exerted as negative refraction can also occur without having a negative refractive index[29]. Aspect (ii) has been investigated by directly measuring the phase and the group propagation time upon propagation of a Gaussian light pulse through a thin metamaterial layer in an interferometric arrangement[19,20]. Notably, a negative ("backward") phase velocity can be accompanied by positive or negative group velocity; whereas, the Poynting vector or the energy velocity is positive ("forward")[19,20] in either case (see the tutorial box).

*Optical metamaterials go three-dimensional*

The obvious next step[30] is to stack an integer number $N>1$ of functional layers of the double-fishnet ($2N+1$ actual layers) to eventually achieve a bulk 3D uniaxial negative-index metamaterials (see Fig. 2(a)). Table 1 summarizes results of corresponding experiments. Even the largest stack with $N=10$ achieved in 2008 still corresponds to a total thickness of the stack significantly smaller than the operation wavelength. This raises the question: How many layers are sufficient to qualify as "3D bulk," *i.e.*, to qualify for convergence when going from a single layer to many layers? Nobody expects the optical properties of a few-monolayer-thin $SiO_2$ film to be significantly different from those of a 50-nm film—despite the fact that the total thickness is much smaller than one optical wavelength in both cases (the situation would be different though for graphene). In metamaterials, however, the situation is more complex, because the building blocks can interact strongly. Thus, one must study how the effective optical parameters change as the number of layers increases. Thus, there is no unique answer to the above question. In some cases, a single functional layer may already be sufficient; in other cases, a few layers are needed. The answer strongly depends on the coupling between adjacent layers; hence, on their spacing. If the distance between adjacent functional layers is small, the coupling is strong, the convergence of the optical properties is slow, and one needs at least four functional layers[34]. The behavior converges, but it does not converge to that of the monolayer case. In contrast, for larger distances, the coupling is weak and the retrieved optical parameters for several functional layers are very closely similar to that of the monolayer case[34]. The retrieved effective parameters remain the same when increasing the number of layers, and we have reached the bulk metamaterial limit. The minimum number of layers qualifying as bulk depends on the metamaterials structure under investigation as well as on the choice of $a_z$. In addition, several groups fabricated multi-layers SRRs, which give negative $\mu$ at THz[12,36] and optical frequencies[35].

Table 1: Summary of fabricated double-fishnet negative-index metamaterials composed of an integer, *N*, of functional layers (or lattice constants) following a corresponding theoretical suggestion [28]. The structure is illustrated in Fig. 2 (a).

| Number of functional layers | Ref. | Year | Operation frequency |
|---|---|---|---|
| *N*=3 | [31] | 2007 | 214 THz |
| *N*=10 | [28] | 2008 | 166 THz |
| *N*=5 | [32] | 2008 | 188 THz |
| *N*=4 | [33] | 2008 | 1 THz |
| *N*=3 | [24] | 2011 | 439 THz |

### *The unit cell goes three-dimensional*

The charm of the double-fishnet structure in Fig. 2(a) is that it is scientifically interesting and fairly simple to fabricate at the same time—even for several functional layers. Standard electron-beam lithography[28,32], focused-ion-beam lithography[24,28], interference lithography[14], and nanoimprint lithography[37] have successfully been used for double-fishnet designs, as well as for other metamaterial structures. However, there is more to three-dimensional metamaterials than just making them thicker.

Flexibility in tailoring the metamaterial's unit-cell interior is another crucial factor, especially if one aims at achieving functionalities other than negative refractive indices. For example, strong chirality—the prerequisite for large optical activity and circular dichroism—requires the lack of a mirror plane parallel to the substrate (for light impinging normal to the surface), not possible with simple layered structures.

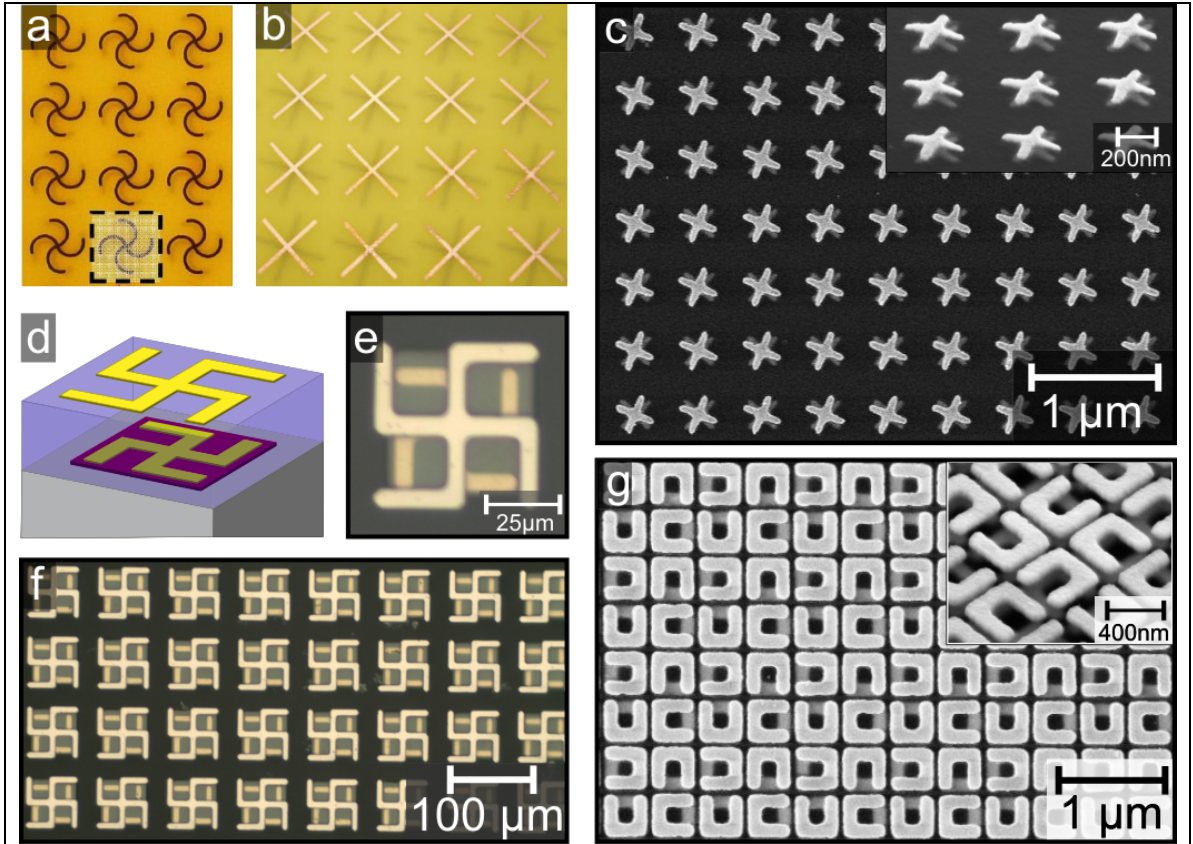

**Figure 3.** (*a*) The experimentally fabricated rosette chiral metamaterials that give negative *n* at GHz frequencies [40]. (*b*) The experimentally fabricated cross-wires chiral metamaterials that give *n*<0 at GHz frequencies [41]. (*c*) Scheme of the chiral structure composed of right-handed twisted gold crosses. Oblique-view electron micrograph of a fabricated sample and normal-view large-area electron micrograph of a right-handed structure [38]. (*d*) Schematic of the unit cell of tunable chiral metamaterial. (*e*) Top view of one unit cell. (*f*) Top-view image of tunable chiral metamaterial on a larger scale [45]. (*g*) Electron micrographs of the fabricated four U's SRRs [39].

*Stacked electron-beam lithography:* Figure 2(b) shows a structure (chiral or multi-layer SRRs) accessible *via* a more general stacking approach that has been followed by several groups[12,35-39,44-45]. Here, a first layer of unit cells, *e.g.*, composed of split-ring resonators, is fabricated by electron-beam lithography and subsequently planarization. For example, the latter can be accomplished by means of a rather thick commercially-available spin-on dielectric, which is then etched to a desired spacer thickness, *e.g.*, *via* reactive-ion etching. The next layer of unit cells is again fabricated using electron-beam lithography aligned with respect to the first layer, etc. Lateral alignments between the different layers down to the 10-nm level have been achieved. While the number of functional layers accessible along these lines is not conceptually limited, the structures published so far have actually been restricted to a few functional layers only, with a maximum of five[32].

Chiral structures are readily accessible using this approach and have led to huge circular dichroism and optical activity at optical frequencies[38,39], but not yet to negative refractive indices—in contrast to microwave[40-42] and far-infrared frequencies[43-45]. In Fig. 3, we summarize the progress of chiral metamaterials[38-45] using this fabrication technique. For the first time, tunable chiral metamaterials were experimentally fabricated, and exhibit tunable optical activity and a tunable refractive index[45].

*Direct laser writing:* The uniaxial gold-helix metamaterial shown in Fig. 2(c) is another paradigm example of a chiral structure[46]. Like the planar SRR, 3D metal helices can be viewed as miniature electromagnetic coils into which the light field can induce an Ohmic current, giving rise to local magnetic-dipole moments. Such structures have not been fabricated by any of the above approaches and likely never will. They are accessible though through direct laser writing (DLW), the three-dimensional counterpart of two-dimensional electron-beam lithography. In regular DLW, femtosecond laser pulses are very tightly focused into the volume of the photo resist. By using two-photon absorption, only a tiny volume is sufficiently exposed by the light. By computer-controlled scanning of the relative positions of focus and resist *via* piezoelectric actuators, almost arbitrary polymer structures can be fabricated[47-49]. In regular DLW, the lateral resolution is limited to about 100 nm. This value obviously cannot compete with state-of-the-art electron-beam lithography, where 10 nm are readily accessible. However, more recent work using stimulated-emission depletion DLW[50] has approached 50-nm lateral resolutions with potential for future improvements. The polymer structures resulting from DLW lithography can be filled with gold using electroplating[46] (also see Fig.2(h)). An electroplating setup can be extremely simple and inexpensive. One merely applies a bias voltage between a transparent electrode on the substrate and a macroscopic counter electrode within a beaker. Thus, accessible metamaterial sample footprints and heights are limited only by the DLW lithography process. For large-scale bulk samples, DLW could be replaced by 3D interference lithography. However, to our knowledge this has not been achieved, yet.

*Other templating approaches*: For other metamaterial structures, anodized alumina templates can be filled with silver or gold, again using electroplating[46] (see Fig. 2(d)). While achieving isotropy remains to be a goal for some of the above metamaterial structures, indefinite or hyperbolic metamaterials rely on intentional anisotropy[51]. Another experimental way to obtain negative refraction[52,53] and focusing is to use metal-dielectric layered metamaterial (see Fig. 2(e)). For electric fields parallel to the metal wire axis in Fig. 2(d) or parallel to the metallic layers in Fig. 2(e), and for frequencies below the effective plasma frequency of the composite, the metamaterial's effective electric permittivity is negative; whereas, it can be positive for the orthogonal polarization of light. The notion "hyperbolic metamaterial" stems from the resulting hyperbolic shape of the iso-frequency surfaces in wave-vector space[51]. These anisotropic systems can be used to achieve broadband all-angle negative refraction and super lens imaging[51-58], for example. Corresponding experiments (negative refraction and sub-wavelength imaging) have been published at microwave[54], infrared[52,55], and visible wavelengths[57].

*Membrane projection lithography*: Yet, more complex designs as shown in Fig. 2(f) can be realized using the directional evaporation technique or "membrane projection lithography." This approach offers a path towards fabricating 3D metamaterials with micrometer scale characteristic dimensions with arbitrary cavity shape and the ability to orient the SRR inclusions along each of the coordinate axes—a crucial step towards creating isotropic metamaterials. Along these lines, the team at Sandia National Laboratories has demonstrated[59] magnetic coupling to the vertically-oriented SRRs at 10-µm-operation wavelength. Their experimental results are in good agreement with the theoretical predictions of the corresponding electromagnetic response. This novel technology is extremely flexible, but also technologically demanding. While only metamaterial structures with a single functional layer have been demonstrated so far, the approach conceptually allows for stacking of layers, just like in electron-beam lithography as discussed above.

*Dielectric three-dimensional optical metamaterials*

All of the above metamaterials involve metals leading to large inherent losses (also see next section), especially at optical or even visible frequencies. Thus, it is interesting to ask whether purely dielectric off-resonant hence low-loss constitutive materials might offer effective magnetic and/or negative-index metamaterials properties as well. Similar for the metallic structures, Mie resonances of dielectric particles play a key role[62] to achieve a negative magnetic permeability $\mu$ [63-69]. It is known from Mie resonance theory that the first resonance of a dielectric sphere is a magnetic-dipole mode (the second resonance for a cylinder). Its wavelength is controlled by the size and the refractive index, $n_s$, of the sphere with diameter, $d$. Precisely, the free-space wavelength, $\lambda_0$, of the first magnetic resonance is given by the condition, $d\, n_s/\lambda_0=(m+1)/2$ with $m=1$, the first electric-dipole resonance by $m=2$. To be used as a sub-wavelength building block in a metamaterials, the wavelength must be larger than the sphere diameter, e.g., $\lambda_0/d >10$. To fulfill this condition, the refractive index of the sphere $n_s$ must be 10 or larger. Two sets of high-dielectric spheres having the same dielectric material, but different radii, or having the same radius of the two spheres and different refractive indices for the two spheres allow for further design freedom. Such a structure is illustrated in Fig. 2(j). For example, one set of spheres can offer electric-dipole moments (leading to negative ε) in the metamaterials, the other set can provide magnetic-dipole moments (negative μ). Due to the low losses of dielectrics, the Mie resonances are very narrow, making the overlapping of the two sets of resonances difficult. Apart from this aspect, three-dimensional isotropic (!) negative-index metamaterials can be realized along these lines. Experiments have been performed at GHz[69] frequencies, where a sphere's diameter is on the order of millimeters. In contrast, at THz or even optical frequencies, no experiments have been published, yet.

Several alternatives arise for 3D isotropic negative-index structures. First, one can combine the magnetic-dipole resonances of dielectric spheres with off-resonant metal wires (see above) to achieve $n<0$[70-71]. Second, high-dielectric spheres embedded in a negative-permittivity plasmonic host, such as metals and semiconductors[72], can be used. Third, high-dielectric spheres for negative $\mu$ can be coated with thin layers of a Drude metal providing negative ε, together leading to $n<0$ at infrared frequencies[73-75]. Fourth, a 3D ensemble of polaritonic spheres for negative $\varepsilon$ was proposed for achieving isotropic $\mu<0$ near the first Mie resonance at infrared frequencies[76-79]. Interestingly, the magnetic-dipole and electric-dipole resonances of high-dielectric spheres can survive in a random or non-periodic configuration[80], enabling low-loss three-dimensional isotropic negative-index photonic metamaterials at infrared and optical frequencies.

*Active loss compensation*

Upon moving from single metal-based metamaterial layers via several layers towards bulk, the metamaterial losses become an increasingly important issue. If, for example, the transmittance of a single metamaterial layer is as large as 90%, the transmittance of hundred layers is a mere $(0.9)^{100} \approx 3 \times 10^{-5}$, which renders the metamaterial essentially opaque, hence useless. The level of losses can be quantified by the so-called figure of merit FOM. This dimensionless number allows for comparing values of metamaterials operating in very different wavelength regimes in a meaningful manner. Mathematically, the FOM is connected to the complex valued refractive index $n$ by FOM=|Re($n$)/Im($n$)|. Intuitively, over a length corresponding to one medium wavelength $\lambda$, the wave amplitude decays to exp(-2$\pi$/FOM). Equivalently, the 1/e intensity decay length is given by FOM $\times \lambda /(4\pi)$. The best-measured values[24,31] on passive double-fishnet negative-index metamaterials are about FOM=3, *i.e.*, the 1/e intensity decay length is just about a quarter of a medium wavelength. These large losses are due to the fact that the region of $n<0$ is very near to the resonance of the single unit cell where Im($n$) is large. Thus, Im($n$) and hence losses can be reduced by moving $n<0$ further away from resonance. This can be accomplished in both the weakly and strongly coupled double-fishnet structures by introducing periodicity effects[34]. While future design optimization might lead to certain improvements, an actually desired increase by several orders of magnitude appears to be out of reach at optical frequencies.

An obvious approach to compensate for the loss is to introduce gain materials into the metamaterials structure. However, a FOM=3 at 1-µm wavelength corresponds to an absorption coefficient of $\alpha \approx 4\times 10^4$ cm$^{-1}$. Gain coefficients of this magnitude are hard to obtain in practice[4]. However, recent experiments[81] on single layers of double-fishnet negative-index structures, in which the intermediate dielectric spacer layer was doped with dye molecules that were pumped by optical picosecond pulses, have raised hopes that this challenge may be hard to solve but not impossible. Corresponding theory[82,83] has been published as well. Other experiments using semiconductor gain in split-ring-resonator metamaterials have only achieved partial loss compensation[84,85]. In principle, under stable steady-state conditions, the FOM can approach infinity at a single wavelength[86,87].

*Conclusions and outlook*

Throughout the last decade, electromagnetic metamaterials have come a long way from microwave frequencies to the visible. More recently, they have also become truly bulk 3D "materials" at optical frequencies, albeit only for certain propagation directions and/or for certain polarizations of light so far. Furthermore, none of these intricate 3D structures is yet available in gram quantities. Thus, one of the future challenges is to fabricate large-scale 3D isotropic metamaterials. As outlined above, some nanofabrication technologies have specifically been developed for this purpose or have been modified accordingly. This comprises stacked electron-beam lithography, membrane projection lithography, direct laser writing and electroplating, as well as particular bottom-up self-assembly approaches[61,88,89] (see, *e.g.*, Fig.2(i)).

Much of the photonic-metamaterial research has been inspired by the fascinating and far-reaching vision of the "perfect lens," as introduced by John Pendry[90], based on the dream of lossless isotropic negative-index metamaterials. While this dream may never come to fruition, it has fueled an entire field. Meanwhile, some researchers have also identified more short-term applications that use essentially two-dimensional metamaterial structures. For example, even meta-surfaces can approach perfect absorbers, *i.e.*, structures that neither transmit nor reflect light in a certain frequency regime and for a broad range of angles[91-96]. Such compact perfect absorbers might prove useful for detectors or energy converters. Other researchers have explored field-enhancement effects for improving the performance of solar cells[97,98]. The magnetic response is also a prerequisite for huge chiral optical effects in three-dimensional metamaterials, *e.g.*, enabling compact broadband circular polarizers[46]. At THz operation frequencies, by using an array of conjugated bi-layer metal resonators with photo-excitable semiconductor inclusions, even actively tunable optical activity with high transmission and very low polarization distortion has been demonstrated experimentally[45]. Other more near-term applications employ the (sharp) metamaterial resonances for sensing purposes[99] *via* their dependence on environment or investigate nonlinear frequency conversion[6,100-106]. In addition, the metamaterial analogue of electromagnetically-induced transparency (EIT)[107-110] shows a transparency window with extremely low absorption and strong adjustable dispersion. The latter could lead to "slow-light" applications from the microwave regime up to THz frequencies, where the structures can be fabricated quite easily. At optical frequencies, however, Ohmic losses do impose certain fundamental constraints.

Finally, when asking for applications or products, one shouldn't forget the field of electromagnetic metamaterials is still rather young. After all, the first "photonic" magnetic metamaterial structure emerged in November 2004—approximately six years ago. Since then, the conceptual, as well as technological, progress has truly been dramatic. The ultimate killer application solving one of the grand challenges still needs to be identified. Metamaterials for improving solar-energy harvesting or metamaterials for novel medical diagnostics are prime candidates.

*Acknowledgments:* We thank Manuel Decker, Jiangfeng Zhou and Thomas Koschny for preparing the figures and useful discussions. Supported by the European Union FET project PHOME (Contract # 213390); by Ames Laboratory, Department of Energy (Basic Energy Sciences) under contract DE-AC02-07CH11358; by the U. S. Office of Naval Research (ONR) under grant No. N000141010925, AFOSR-MURI under grant No. FA9550-06-1-0337, and by the European Union project NIM_NIL (Contract # 228637) (CMS); by "Deutsche Forschungsgemeinschaft" through subprojects CFN A1.4 and A1.5, and by "Bundesministerium für Bildung und Forschung" through the project METAMAT (MW).

*Tutorial box*

The notion "negative index of refraction" has given rise to plenty of confusion. One aspect is that it needs to be specified whether one means a negative *phase* index, $n_p$, or a negative *group* index, $n_g$. In either case, "negative" refers to the direction of the phase or group velocity vector with respect to the electromagnetic energy flow (Poynting vector), which is tacitly assumed to be positive or forward. Fig.B1 illustrates these quantities for the simple case of an isotropic effective material with $\varepsilon(\omega)=\mu(\omega)$, such that no reflections occur at the interface of the material and vacuum or air.

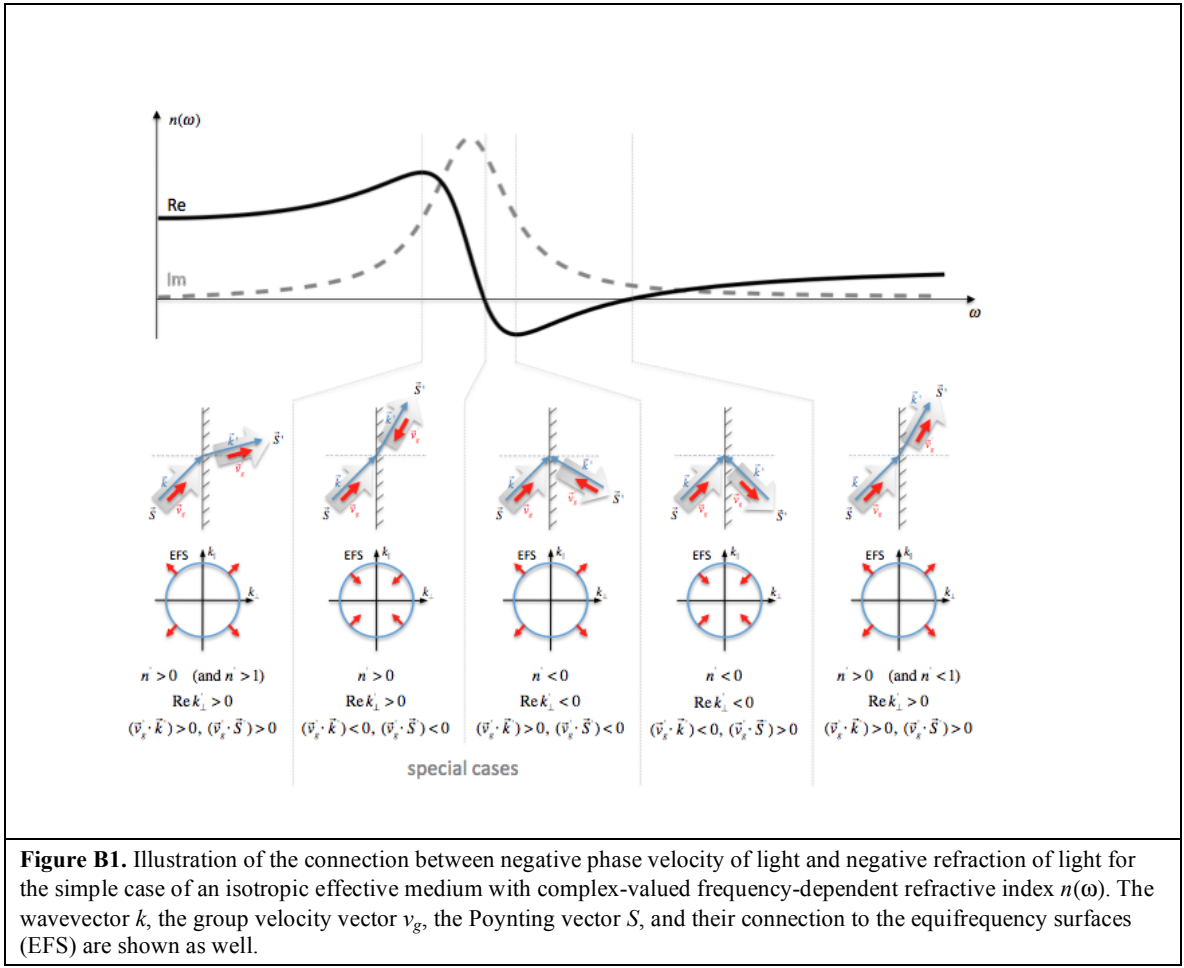

**Figure B1.** Illustration of the connection between negative phase velocity of light and negative refraction of light for the simple case of an isotropic effective medium with complex-valued frequency-dependent refractive index $n(\omega)$. The wavevector $k$, the group velocity vector $v_g$, the Poynting vector $S$, and their connection to the equifrequency surfaces (EFS) are shown as well.

On the top of Fig. B1, we plot the frequency dependence of the isotropic index of refraction $n(\omega)$. In the middle of the figure B1, we plot the wavevector $k$, the group velocity vector $v_g$, the Poynting vector $S$ for the interface of air and the dispersive isotropic metamaterial. In addition, we plot the equifrequency surfaces (EFS) for the dispersion relation $n(\omega)$. If the equifrequency contours move outwards with increasing frequency then $\vec{v}_g \cdot \vec{k} > 0$; if they move inwards $\vec{v}_g \cdot \vec{k} < 0$. For any general case $\vec{v}_p = (c/|n_p|)\hat{k}$ with $\hat{k} = \vec{k}/k$. We also have $\vec{v}_g = \nabla_k \omega = (c/|n_g|)\hat{k}$, where $n_g = \omega d|n_p|/d\omega + |n_p|$. In addition, we would like to mention that $|\vec{k}| = \sqrt{k_x^2 + k_y^2} = R(\omega)$, and $|\vec{v}_p| = \omega/|k| = c/|n_p|$, $\omega = cR(\omega)/|n_p|$ where $R$ is the radius of the circular EFS.

The "special cases" require some discussion. Because of the anomalous dispersion in these regions, the energy flow, given by the Poynting vector S, and the group velocity point in opposite directions. Causality still enforces energy flow to the right, away from the interface. This counter direction of group velocity and energy flow is only possible because of the strong absorption accompanying the anomalous dispersion region which causes an incident pulse to eventually broaden and diminish in amplitude away from the interface. Within the anomalous dispersion region, the phase velocity may be either parallel or anti-parallel to the energy flow or group velocity resulting in either positive or negative refraction of light. The anomalous-dispersion region exists for arbitrarily small yet finite losses and only disappears for the pathological case of strictly zero loss. The negative group velocity corresponds to "superluminal" pulse transmission, *i.e.*, the center of a Gaussian pulse appears on the rear side of a slab before it enters its front side. This mind-boggling behavior is, however, not connected to superluminal energy transfer due to the energy that gets stuck inside the material *via* the inherent losses in this spectral region.

A negative *group* index is nothing special at all. It can appear in the region of anomalous dispersion of any normal dielectric resonance (*i.e.*, $\mu=1$) or for many dielectric photonic crystals. A negative group index can lead to negative refraction of light at an interface with air – even if the phase index is positive. Confusingly, a negative phase index also leads to negative refraction of light at an interface with air – even if the group index is positive. Negative refraction can also occur for birefringent dielectric materials with only positive phase and positive group indices. This confusion originates from the fact that Snell's law of refraction makes a statement on the angle of refraction of rays in geometrical optics, or on the direction of the Poynting vector in wave optics. However, Snell's law generally makes no direct statement on the direction of the phase velocity vector. Thus, one must generally be very cautious when interpreting negative-refraction experiments.

The *phase* refractive index $n$ is simply defined via $c=c_0/n$, where $c$ is the phase velocity of light and $c_0$ the vacuum speed of light. Thus, negative $n$ immediately leads to negative phase velocity, which means that the optical path length becomes negative. Hence, the (phase) propagation time through a slab of a structure with thickness $d$ becomes negative as well. A negative phase propagation time, $\Delta t<0$, can be measured in direct interferometric experiments. However, caution must be exerted when interpreting such experiments. Mathematically, a phase refractive index can directly be inferred via $n=c_0\Delta t/d$ (upon appropriately separating effects due to the slab's surfaces). Yet, if the derived quantity $n$ is actually a *material* property, it must only depend on the type of material and not on its thickness $d$. Thus, for example, if one doubles the slab's thickness, $\Delta t$ must obviously double as well, leading to the identical value for $n$. We call a structure a "bulk metamaterial" if this condition is met (usually not the case for very thin metamaterial structures). The convergence from a single layer towards bulk needs to be investigated for each metamaterial structure separately (see main text).

Finally, we note that much of our above simplified discussion has tacitly assumed that the optical properties of the (meta)material under discussion are isotropic, which is only very rarely fulfilled in experiments. If not, *the* refractive index has to be replaced by the appropriate tensor element.